\documentclass[12 pt]{article}
\usepackage{epsf}
\newcommand{\bee}{\begin{equation}}
\newcommand{\ee}{\end{equation}}
\newcommand{\bea}{\begin{eqnarray}}
\newcommand{\eea}{\end{eqnarray}}
\newcommand{\R}{\rm I\kern-.2emR}
\newcommand{\C}{\rm \kern.25em\vrule height1.4ex
depth-.12ex width.06em\kern-.31em C}
\newcommand{\N}{{\rm I\kern-.16em N}}
\newcommand{\Z}{{\rm Z\kern-.35em Z}}

\begin{document}                                                                
\thispagestyle{empty}                                                           
\begin{flushright}                                                              
MPI-PhT/2000-45\\
AZPH-TH/00-03\\
November 2000
\end{flushright}                                                                
\bigskip\bigskip                                                                
\begin{center}                                                                  
{\Huge Percolation and the existence of a soft}\\
\vglue2mm
{\Huge phase in the classical Heisenberg model}
\end{center}                                                                    
\vskip 1.0truecm                                                                
\centerline{\bf                                                                 
Adrian Patrascioiu}                                                             
\vskip5mm                                                                       
\centerline{Physics Department, University of Arizona}                          
\centerline{Tucson, AZ 85721, U.S.A.}                                           
\vskip5mm                                                                       
\centerline{and}                                                               
\vskip5mm                                                                       
\centerline{\bf Erhard Seiler}                                                 
\vskip5mm                                                                     
\centerline{Max-Planck-Institut f\"{u}r                                        
 Physik}                                                                       
\centerline{ -- Werner-Heisenberg-Institut -- }                                
\centerline{F\"ohringer Ring 6, 80805 Munich, Germany}                         
\vskip 2cm                                                                      
\bigskip \nopagebreak                                                           
\begin{abstract}
\noindent
We present the results of a numerical investigation of percolation
properties in a version of the classical Heisenberg model. In particular 
we study the percolation properties of the subsets of the lattice
corresponding to equatorial strips of the target manifold ${\cal S}^2$.
As shown by us several years ago, this is relevant for the existence 
of a massless phase of the model. Our investigation yields strong
evidence that such a massless phase does indeed exits. It is further
shown that this result implies lack of asymptotic freedom in the 
massive continuum limit. A heuristic estimate of the transition 
temperature is given which is consistent with the numerical data.
\end{abstract}

\section{Introduction}

If one looks at textbooks to learn about the phase diagram of
the two dimensional ($2D$) $O(N)$ models the situation seems clear: 
for $N=2$, there is a transition to a low temperature phase with only
power law decay of correlations, whereas for the nonabelian case $N>2$
there is exponential decay at all temperatures. But while the first
statement has been proven rigorously a long time ago \cite{fs}, the
second one remains an open mathematical question \cite{www}. The
standard belief is rooted in the perturbative asymptotic freedom of the 
models for $N>2$; but over the years we have brought forth many reasons 
why we think it is unfounded \cite{apz,ff,super}. The absence of a 
mathematical proof together with ambiguous numerical results left
the issue wide open.

In this paper we would like to present what we regard as convincing
numerical evidence that in fact the $2D$ $O(3)$ model possesses a 
massless phase for sufficiently large $\beta$ and give a rigorous proof
that this is incompatible with asymptotic freedom in the massive phase.
We will also give a heuristic explanation of why and where the phase 
transition happens.

The models we are considering consist of classical spins $s$ taking 
values on the unit sphere $S^{N-1}$, placed at the sites of a $2D$
regular lattice. These spins interact ferromagnetically with their 
nearest neighbors. Let $\langle ij \rangle$ denote a pair of neighboring 
sites. We will consider two types of interactions between neighbouring 
spins:
\begin{itemize}
\item Standard action (s.a.):  $H_{ij}=-s(i)\cdot s(j)$
\item Constrained action (c.a.):  $H_{ij}=-s(i)\cdot s(j)$
for $s(i)\cdot s(j) \geq c$
and
$H_{ij}=\infty$ for $s(i)\cdot s(j)<c$ for some $c\in [-1,1)$.
\end{itemize}
The corresponding Gibbs measures are (for a finite lattice) given by
\bee
d\mu_{s.a.}={1\over Z} \prod_{\langle i j\rangle}
e^{-\beta H_{ij}} \prod_i d\nu\left(s(i)\right)
\ee
for the standard action and
\bee
d\mu_{c.a.}={1\over Z} \prod_{\langle i j\rangle}
\left[e^{-\beta H_{ij}} \theta\left(s(i)\cdot s(j)-c\right)\right]
\prod_i d\nu\left(s(i)\right)
\ee
for the constrained action, where $d\nu$ is the standard measure 
on the two sphere $S^2$ and the product $\prod_{\langle i j\rangle}$ 
is over nearest neighbors.

Almost a decade ago we showed \cite{ael} that one can rephrase the 
question of the existence of a soft phase in these models as a
percolation problem and in fact this is the reason we introduced the c.a.
model. It should be noted that the c.a. model shares with the s.a.
model not only invariance under $O(N)$, but has also the same 
perturbative (= low temperature) expansion and the same `smooth' 
classical solutions as the s.a. model. It is therefore to be expected 
that the s.a. and c.a.
models fall in the same universality class (possess the same continuum
limit) and, as we shall show shortly, the numerical evidence supports
this expectation. The advantage of studying the c.a. model stems from
the following fact: let $\epsilon_c=\sqrt{2(1-c)}$ and $S_{\epsilon_c}$ 
the set of sites such that $|s\cdot n|<\epsilon_c/2$ for some given 
unit vector $n$. Our rigorous result \cite{ael} was that if for some
$\epsilon>\epsilon_c$ the set $S_{\epsilon}$ on the triangular (T) 
lattice does not contain a percolating cluster, then the
$O(N)$ model must be massless at that $c$. For the abelian $O(2)$
model we could prove the absence of percolation of this equatorial set
$S_{\epsilon_c}$ for $c$ sufficiently large \cite{ael} (modulo certain
technical assumptions which were later eliminated by  M. Aizenman 
\cite{aiz}). For the nonabelian cases we could not give
a rigorous proof. We did however present certain arguments \cite{ap,npb} 
explaining why the percolating scenario seemed unlikely.

In this paper we will present numerical evidence that there is
an $\epsilon_)$ such that for $\epsilon\leq \epsilon_0$
$S_{\epsilon}$ does not percolate for any $c$;
for sufficiently large $c$ this $\epsilon_0$ will be larger than
$\epsilon_c$ and the model will thus be massless. We will also show 
that due to a rigorous inequality derived by us in the past \cite{conf},
the existence of a finite $\beta_{crt}$ in the s.a. model on the square 
(S) lattice is incompatible with the presence of asymptotic freedom in 
the massive continuum limit of the model.
\section {Percolation and masslessness}
In this section we briefly review the special features of percolation
in two dimensions and give a brief sketch of our argument relating
percolation properties to the absence of a mass gap. We restrict the
discussion to the T lattice; this keeps the arguments simpler because
the T lattice is self-matching and no distinction has to be made
between connectedness and $\star$-connectedness (where points are also
considered connected along diagonals).

The following two facts special to $2D$ are relevant for our
discussion:

1. {\it Noncoexistence of disjoint percolating sets}:
Let $A$ be the subset of the lattice defined by the spin
lying in some subset ${\cal A}\subset S^{N-1}$ and $\tilde A$ its 
complement. Then with probability 1 $A$ and $\tilde A$ do not 
percolate at the same time. This has been proven rigorously
only for special cases like Bernoulli percolation and the $+$ and $-$ 
clusters of the Ising model, but is believed to hold quite generally.
(Aizenman \cite{aiz} showed that in the case of $O(2)$ one does not
need to invoke this principle).

2. {\it Russo's lemma} \cite{russo}:
If neither $A$ nor its complement $\tilde A$ percolate,
then the expected size of the cluster of $A$ attached to the origin,
denoted by $\langle A\rangle$, diverges; the same holds for its
complement $\tilde A$. 
(In this simple form the lemma only holds for a self matching lattice
like the T lattice).
If $\tilde A$ percolates, then $\langle A\rangle$ is expected to be
finite.

The subsets of the sphere $S^2$ that interest us here are the following:
\begin{itemize}
\item
`equatorial strip' ${\cal S}_\epsilon$, defined by
$|s\cdot n|<\epsilon/2$ for some fixed unit vector $n$.
\item
`upper polar cap' ${\cal P}_\epsilon^+$, defined by
$s\cdot n\geq\epsilon/2$,
\item
`lower polar cap' ${\cal P}_\epsilon^-$, defined by
$s\cdot n\leq -\epsilon/2$.
\item `union of polar caps' ${\cal P}_\epsilon = {\cal P}_\epsilon^+\cup
{\cal P}_\epsilon^-$.
\end{itemize}
The subsets of the lattice defined	by these subsets of the sphere
we denote by the corresponding roman letters $S_\epsilon$ etc. and
for brevity we say `a certain subset of the sphere percolates'
instead of `the subset of the lattice induced by a certain subset
of the sphere percolates' etc..

According to the discussion above, there are the following possibilities: either 
$S_\epsilon$ percolates, or $P_\epsilon$ percolates, or neither 
$S_\epsilon$ nor $P_\epsilon$ percolates and then both have divergent 
mean size (we shall call this third possibility in short {\it formation 
of rings}). 

Let us now briefly review our argument \cite{ael} that relates 
percolation properties to the absence of a mass gap. Our statement was 
that if there was an equatorial strip ${\cal S}_\epsilon$ that did not
percolate for a certain $c>1-\epsilon^2/2$, there could be no mass gap
in the system.

The argument is based on the imbedded Ising variables 
$\sigma_i\equiv {\rm sgn}(s(i))$. Using these variables, the s.a.
Hamiltonian becomes:
\bee
H_{ij}=-\sigma_i\sigma_j |s_\|(i) s_\|(j)|-s_\perp(i)\cdot s_\perp(j)
\ee
where $s_\|(i)=s(i)\cdot n$ and $s_\perp(i)=n\times (s(i)\times n)$.
The c.a. model can be similarly described in terms of the variables
$\sigma_i$, $|s_\|(i)|$ and $s_\perp(i)$. In both models one thus
obtains an induced Ising model for which the Fortuin-Kastleyn (FK)
representation \cite{fk} is applicable. In this representation the Ising 
system is mapped into a bond percolation problem: In the s.a. model
a bond is placed between any like neighboring Ising spins with 
probability $p=1-\exp(-2\beta s_\|(i)s_\|(j))$. For the c.a. model a 
bond is also placed if after flipping one of the two neighboring Ising 
spins the constraint $s(i)\cdot s(j) \geq c$ is violated. From the FK
representation is follows that the mean cluster size of the site 
clusters joined by occupied bonds (FK-clusters) is equal to the Ising 
magnetic susceptibility. In a massive phase the latter must remain 
finite. Hence, if the FK-clusters have divergent mean size, the 
original $O(3)$ ferromagnet must be massless (the Ising variables 
$\sigma$ are local functions of the originally spin variables $s$).

Now notice that by construction for the c.a. model the FK-clusters with,
say, $\sigma=+1$ must contain all sites with $s(i)\cdot n>\sqrt{(1-c)/2}$.
Therefore the model must be massless if clusters obeying this condition
have divergent mean size. But the polar set $P_\epsilon$ consists
of the two disjoint components $P^+_\epsilon$ and $P^-_\epsilon$.
For $c>1-\epsilon^2/2$ there are no clusters containing
elements of both $P^+_\epsilon$ and $P^-_\epsilon$. Hence if for such
values of $c$ clusters of $P_\epsilon$ form rings, so do clusters of
$P^+_\epsilon$ separately and the $O(3)$ model must be massless by the
argument just given.

If we want to study the percolation or absence of percolation 
of the set $A$ corresponding to a subset ${\cal A}\subset S^{N-1}$ of
positive measure numerically, we have to consider a sequence of tori 
of increasing size $L$. On these tori we measure the mean cluster size of
of $A$. If $A$ percolates in the thermodynamic limit, by translation 
invariance $\langle A\rangle=O(L^2)$; if its complement percolates
$\langle A\rangle$ should approach a finite nonzero value, and if $A$
forms rings we expect $\langle A\rangle=O(L^{2-\eta})$ for some $\eta>0$.
Therefore, if we define the ratio
\bee
r=\langle P_\epsilon\rangle/\langle S_\epsilon\rangle     ,
\ee
for $L\to\infty$ it should either go to 0
if $S_\epsilon$ percolates or to $\infty$ if $P_\epsilon$ percolates;
if neither $S_\epsilon$ nor $P_\epsilon$ percolates, then both form
rings and the ratio $r$ could diverge, go to 0 or approach some finite,
nonzero value depending upon the value of the critical index $\eta$ for 
the two types of clusters.

In the next section we will describe what our numerical simulations tell
us about the percolation properties of equatorial strips and polar caps.

\section{Main numerical results}

\begin{figure}[htb]
\centerline{\epsfxsize=9.0cm\epsfbox{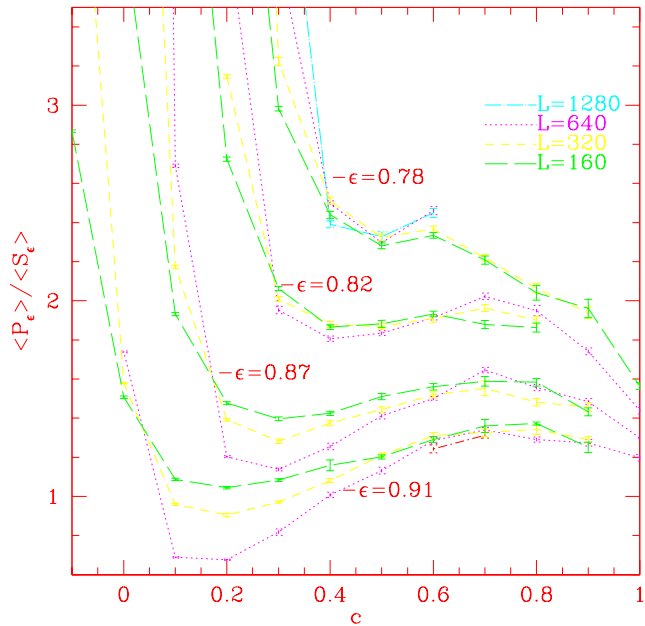}}
\caption{Ratio $\langle{P_\epsilon\rangle/\langle S_\epsilon\rangle}$
for various $\epsilon$ values versus $c$}
\label{tr}
\end{figure}

Our results were obtained from a Monte Carlo (MC) investigation using an 
$O(3)$ version of the Swendsen-Wang cluster algorithm \cite{sw} and
consist of a minimum of 20,000 lattice configurations used for taking
measurements. For each value of $\epsilon$ we studied $L=160$, 320 and 640
(for $\epsilon=.78$ we also studied $L=1280$). 

In Fig.\ref{tr} we show the numerical value of the ratio $r$ as function
of $c$ for $\beta=0$ for four values of $\epsilon$ for the c.a. model on
a T lattice. Three distinct regimes are manifest for each of the four
values of $\epsilon$ investigated: 
\begin{itemize}
\item For small $c$, $r$ is increasing with $L$, presumably diverging
to $\infty$ (region 1).
\item For intermediate $c$, $r$ is decreasing with $L$, presumably
converging to 0 (region 2).
\item For $c$ sufficiently large depending upon $\epsilon$, $r$ 
shows a very mild dependence upon $L$.
\end{itemize}

This can only mean that for these values of $\epsilon$ for small $c$
$P_\epsilon$ percolates, for intermediate $c$ $S_\epsilon$
percolates and for sufficiently large $c$ both $P_\epsilon$ and
$S_\epsilon$ form rings with quite similar (possibly equal)
values of $\eta$. Our data allow to deduce a semiquantitative `phase
diagram' in the $(c,\epsilon)$-plane of the percolation problem induced by
the c.a. model on the T lattice for $\beta=0$. 

\begin{figure}[htb]
\centerline{\epsfxsize=9.0cm\epsfbox{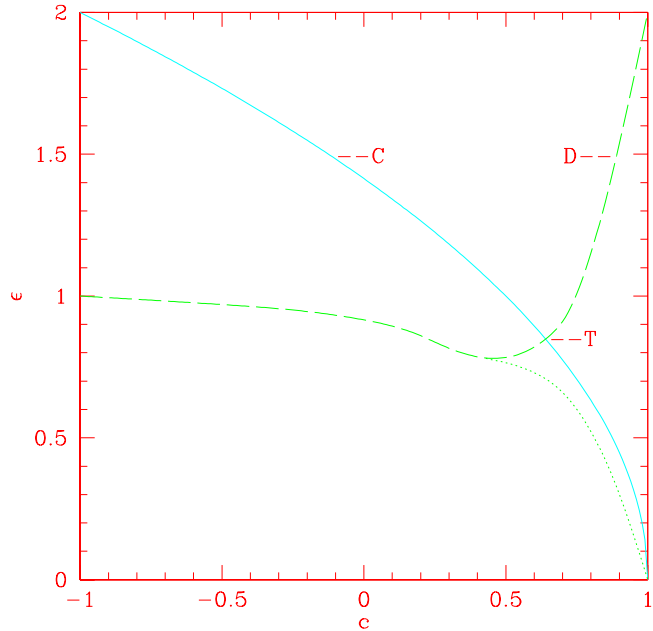}}
\caption{Phase diagram of the $O(3)$ model on the T lattice}
\label{phase}
\end{figure}

This is shown in Fig.\ref{phase}. The solid line C is the curve
$c=1-\epsilon^2/2$; above that line the two polar caps cannot
touch and therefore their union cannot percolate. The dashed line D
represents the minimal equatorial width above which $S_\epsilon$ 
percolates. The point T at the intersection of the curves D and C
gives an upper bound for $c_{crt}$, the value of $c$ above which the 
c.a. model is massless. 

Let us explain how this picture was obtained: since for $c=-1$ (no 
constraint) the model reduces to independent site percolation, for which 
the percolation threshold is known rigorously to be $\epsilon=1$, curve 
D has to start at $\epsilon=1$. With increasing $c$ that threshold
shifts to smaller values of $\epsilon$. Four points of the dashed line
are in fact determined by the data displayed in Fig.\ref{tr}: the clearly
identifiable four points where the lines for different lattice sizes 
$L$ cross and the ratio $r$ becomes independent of $L$ determine the
percolation threshold for the chosen value of $\epsilon$ and so 
determine a point of the dashed line D.

Two features of this diagram are worth emphasizing: \\
1. An equatorial strip of width less than approximately $\epsilon=.76$
{\it never} percolates.\\
2. In Fig.\ref{tr} for approximately $c>0.4$ a range of $\epsilon$'s
appears such that the ratio $r$ again becomes approximately independent 
of $L$. This is signalling the appearance of a new `phase' in which both 
$S_\epsilon$ and $P_\epsilon$ form rings (the dotted line separates it 
from the region of percolation of $P_\epsilon$).

This regime of ring formation of both $S_\epsilon$ and $P_\epsilon$ is
lying between the dotted and the dashed lines in Fig.\ref{phase}. Our
data give strong evidence of its existence, but they do not determine
in detail where the boundaries are. The dotted line has to run to
$\epsilon=0$ for $c=1$ because below it there is percolation of
$P_\epsilon$, and this is not possible above the solid line C
$\epsilon=\epsilon_c$ (because it would conflict with the principle
of non-coexistence of disjoint percolating sets). We drew the dashed 
line into upper right corner because we expect that eventually, for
$c$ approaching 1, any polar cap will start forming rings, thereby
preventing percolation of the corresponding equatorial strips.

But what it is essential for our conclusion that there is a massless
phase is only that there is a regime below the dashed line D and above
the solid line C, in which $S_\epsilon$ does not percolate and the 
two polar caps do not touch. In other words, the lines C and D have
to cross (the crossing point is denoted by T in Fig.\ref{phase}). 
Since we found that for $\epsilon<\epsilon_0=0.76$ ${\cal S}_\epsilon$ 
never percolates, this means that for $c\geq c_{\epsilon_0}= 0.71$ the 
model is massless. In fact the massless phase must start earlier, and 
for instance based on our data we estimate  
that at c=0.61 the c.a. model is already massless.

In the next section we will further corroborate the fact that
for $\epsilon<0.76$ the equatorial strip does not percolate for any $c$.

We would like to comment briefly on another recent paper dealing with
percolation properties of equatorial strips in the $O(3)$ model:
All\`es et al \cite{alles} published a study showing that for 
$\epsilon=1.05$ and $\beta=2.0$ in the s.a. model $S_\epsilon$ 
percolates. Although strictly speaking our percolation argument applies 
only to the c.a. model, the result of All\`es et al is not surprising 
since at $\beta=2.0$ the s.a. model is clearly in its massive phase 
\cite{apost}, hence, by analogy with what happens in the c.a. model, 
one would expect that clusters of a sufficiently wide equatorial strip 
percolate (see \cite{com}). The real
issue, which the authors of \cite{alles} did not seem to appreciate, is
whether in the c.a. model clusters of the equatorial strip 
$S_{\epsilon_c}$ continue to percolate for $c$ sufficiently close to 1. 
The numerics presented in Fig.\ref{tr} suggest 
that that is not the case.
\clearpage
\section{Corroborating numerics}

\begin{figure}[htb]
\centerline{\epsfxsize=9.0cm\epsfbox{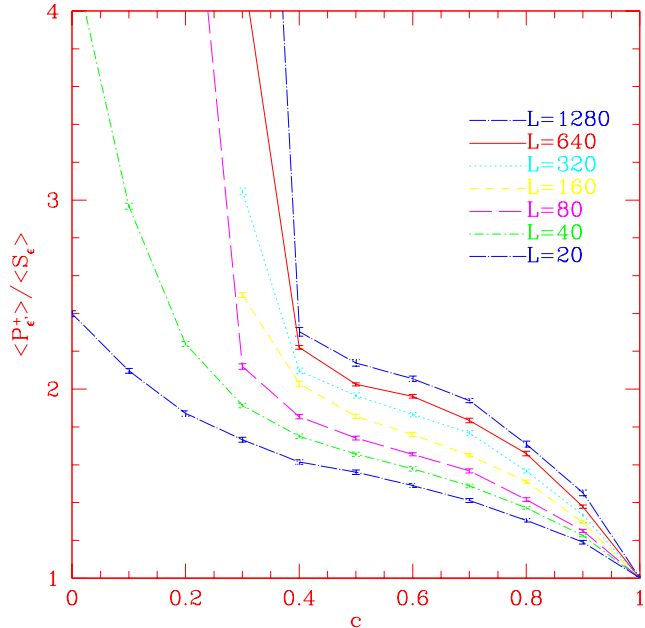}}
\caption{The ratio of the mean cluster size of 
a polar cap of height .75 to that of an equatorial
strip of the same height}
\label{rings}
\end{figure}

To corroborate our most important result, namely that for approximately
$\epsilon<.76$ $S_\epsilon$ does not percolate for any value of $c$,
we also measured (at $\beta=0$) the ratio of the mean cluster size of 
the set $P^+_{\epsilon'}$ with $\epsilon'=.5$ to that of the set 
$S_\epsilon$ with $\epsilon=0.75$ ($\epsilon'$ was chosen so that 
$P^+_{\epsilon'}$ has equal density with $S_\epsilon$).
The results are shown in (Fig.\ref{rings}). This figure shows that
for $c$ less than about 0.4 (and greater than 0) the ratio grows very
rapidly with $L$, indicating that $P^+_{\epsilon'}$ forms rings while 
$S_\epsilon$ has finite mean size; this region terminates around $c=0.4$,
where presumably also $S_\epsilon$ starts forming rings, and the 
dependence of the ratio upon $L$ becomes much milder. Since for $c>0.4$ 
the ratio continues to grow with $L$, at equal density, clusters of the 
polar cap are larger than those of the equatorial strip.
The larger average cluster size of the polar cap compared to the strip 
of the same area is probably due to the fact that the strip has a
larger boundary than the polar cap. 
This is in agreement with a general conjecture stated in
\cite{ap}, namely that for $c$ sufficiently large, if two sets have equal area
but different perimeters, the one with the smaller perimeter will
eventually, for $c$ approaching 1, have larger average cluster size. 
For the case at hand, this is apparently true for all values of $c$.

\clearpage
\section{Comparison to the $O(2)$ model}

\begin{figure}[htb]
\centerline{\epsfxsize=9.0cm\epsfbox{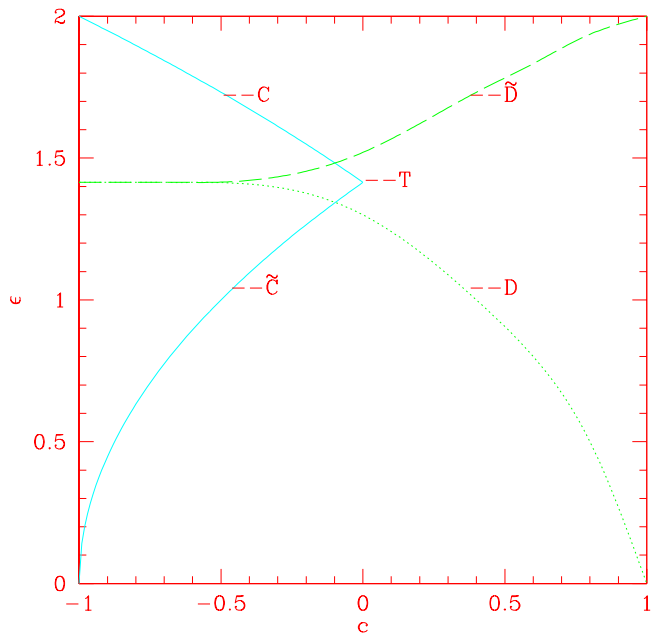}}
\caption{The phase diagram for the $O(2)$ model on the T lattice}
\label{phaseo2}
\end{figure}

The general belief, which we criticized in ref.\cite{ff}, is that there is
a fundamental difference between abelian and nonabelian models. To test 
this belief we also studied the ratio $r$ for the c.a. $O(2)$ model on 
the T lattice. The phase diagram is shown in Fig.\ref{phaseo2}. 
Since in the $O(2)$ model the set ${\cal P}_\epsilon$ can also be 
regarded as a set ${\cal S}_{\tilde \epsilon}$ where 
$\tilde \epsilon=\sqrt{4-\epsilon^2}$, certain features of that diagram
follow from rigorous arguments. For instance it is clear that in the
c.a. model there exist two intersecting curves C and $\tilde{\rm C}$
and in the region to their right the model must be 
massless \cite{ael,aiz}. The precise location of the curves
$D$ (or $\tilde D$) must be determined numerically, 
something which we did not do. We did verify though that the ring 
formation region begins around $c=-0.5$.
 
\section{Universality between the s.a. and c.a. models}

In our opinion the arguments and numerical evidence provided so far give
strong indications that the c.a. $O(3)$ model on the T lattice has
a massless phase. 
Universality would suggest that a similar situation must exist for the 
s.a. models on the T and S lattices. To test universality we measured
on the S lattice the renormalized coupling
both on thermodynamic lattices in the massive phase and in finite volume
in the presumed critical regime (as in \cite{kp}). Our data for the c.a.
model on the S lattice only determine an interval (about $.5$ to $.7$) in
which the
massless phase of the model sets in; we tried to see if we could get a
similar $L$ dependence for the renormalized coupling in the s.a. 
model at a suitable $\beta$ as for $c=.61$ in the c.a. model at
$\beta=0$.
This seems to be indeed the case for $\beta$ roughly $3.4$. We went only
up to $L=640$, hence this equivalence between $c$ and $\beta$ should be
considered only as a rough approximation, but there seems to be no doubt
that the two models have the same continuum limit.

\begin{figure}[htb]
\centerline{\epsfxsize=9.0cm\epsfbox{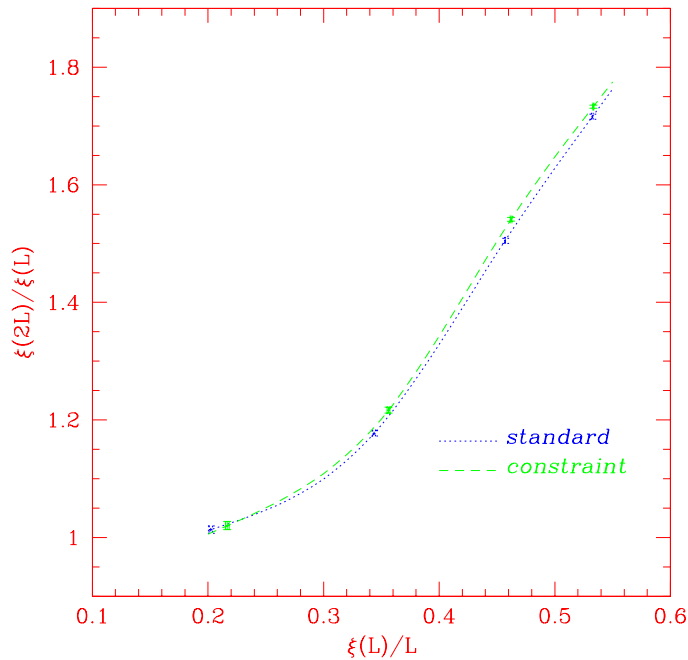}}
\caption{The step scaling curves of the s.a. and c.a.
$O(3)$ models on the T lattice}
\label{scale}
\end{figure}

We also compared the {\it step scaling curve} in the s.a. and
c.a. models. The step scaling curve is obtained as follows: on a 
periodic lattice of size $L\times L$ we define an apparent correlation
length $\xi(L)$. Namely let $P=(p,0)$,
$p={2n\pi\over L}$, $n=0,1,2,...,L-1$. Then define
\bee
  \xi(L)={\sqrt{3}\over 4\sin(\pi/L)}\sqrt{ G(0)/G(1)-1 }
\ee
where
\bee
        G(p)={1\over L^2}\langle |\hat s(P)|^2\rangle;\ \
\hat s(P)=\sum_x e^{iPx} s(x)
\ee
Leaving $\beta$ respectively $c$ fixed, one doubles $L$ and measures
also $\xi(2L)$. The step scaling function gives the ratio 
$2\xi(L)/\xi(2L)$ versus $L/\xi(L)$. In the continuum limit 
($L\to\infty$ and $\beta\to\beta_{\it crt}$ at $L/\xi(L)$ fixed), this 
procedure produces a unique curve characterizing
the universality class of the model. In Fig.\ref{scale} we present the
step scaling function for the c.a. and s.a. models. The data were
produced by adjusting $\beta$ and $c$ so that at $L=20$ we obtain
roughly the same $\xi(L)$ in the two models. After that, leaving 
$\beta$ respectively $c$ fixed, $L$ was doubled until $L=320$. As can be
seen, the two step scaling curves agree reasonably well; the slight
disagreement is probably due to the fact that the two curves have to
agree only in the continuum limit ($\xi\to\infty$), i.e. they have 
different lattice artefacts, whereas the largest value of $\xi$ reached 
was only approximately 35 lattice units.
\section{Heuristic explanation of the transition}

It is intersting to note that there is a heuristic explanation for both
the existence of a massless phase in the s.a. $O(3)$ model and for the 
value of $\beta_{\it crt}$. Indeed it is known rigorously that in $2D$
a continuous symmetry cannot be broken at any finite $\beta$. 
In a previuos paper \cite{super} we 
showed that the dominant configurations at large $\beta$ are not 
instantons but superinstantons (s.i.). In principle both instantons and s.i.
could enforce the $O(3)$ symmetry. In a box of diameter $R$ the 
former have a minimal energy $E_{\it inst}=4\pi$ \cite{bp} while the
latter $E_{s.i.}=\delta^2\pi/\ln R$, where $\delta$ is the angle by which
the spin has rotated over the distance $R$. Now suppose that 
$\beta_{\it crt}$ is sufficiently large for classical configurations to 
be dominant. Then let us choose $\delta=2\pi$ (restoration of symmetry) 
and ask how large must $R$ be so that the superinstanton configuration 
has the same energy as one instanton. One finds $\ln R=\pi^2$. But in the
Gaussian approximation

\bee
\langle s(0)\cdot s(x)\rangle\approx 1-{1\over \beta\pi}\ln x
\ee

Thus restoration of symmetry occurs for $\ln x\approx\pi\beta$. This
simpleminded argument suggests that for $\beta\geq\pi$ instantons become
less important than s.i.. Now in a gas of s.i. the image of any small
patch of the sphere forms rings, hence the system is massless. While this
is not a quantitative argument, we believe it captures qualitatively what
happens: a transition from localized defects (instantons) to 
super-instantons.
\section{Absence of asymptotic freedom}

Next let us discuss the connection between a finite $\beta_{crt}$
and the absence of asymptotic freedom. It follows from our earlier work
concerning the conformal properties of the critical $O(2)$
model \cite{conf}. We refer the reader for details to that paper and give
only an outline of the argument. The s.a. lattice $O(N)$ model possesses
a conserved isospin current. This currrent can be decomposed into a
transverse and longitudinal part. Let $F^T(p)$ and $F^L(p)$ denote the 
thermodynamic values of the 2-point functions of the transverse and
longitudinal parts at momentum $p$, respectively. Using reflection 
positivity and a Ward identity we proved that in the massive continuum 
limit the following inequalities must hold for any $p\neq 0$:
\bee
    0\leq F^T(p)\leq F^T(0)=F^L(0)\leq F^L(p)=2\beta E/N
\ee
Here $E$ is the expectation value of the energy 
$$E=\langle s(i)\cdot s(j)\rangle$$ at inverse temperature $\beta$. Since
$E \leq 1$ it follws that if $\beta_{\it crt}<\infty$ $F^T(0)-F^T(p)$ 
cannot diverge for $p\to\infty$ as required by perturbative asymptotic 
freedom \cite{balog}. In fact, for $\beta_{crt}=3.4$ (which is a 
reasonable guess) $F^T(p)$ must be less than 2.27, in violation of the
form factor computation giving $F^T(0)-F^T(\infty)>3.651$ \cite{bn}.

\section{Concluding remarks}

Since the implications of our result, that for the c.a. model a 
sufficiently narrow equatorial strip never percolates, are so dramatic,
the reader may wonder how credible are the numerics. The only debatable 
point is whether our results represent the true thermodynamic behaviour 
for $L\to\infty$ or are merely small volume artefacts. While we cannot 
rule out rigorously the latter possibility, certain features of the data 
make it highly implausible:

\begin{itemize}
\item Small volume effects should set in gradually, while the data in
Fig.\ref{tr} indicate a rather abrupt change from a region where $r$ is
decreasing with $L$ to one where $r$ is essentially independent of $L$.
\item
For $c\to 1$ at fixed $L$, $r$ must approach the `geometric' value 
$r=2/\epsilon-1$. As can be seen, in all the cases studied, throughout 
the `ring' region $r$ is clearly larger than this value,
while it should go to 0 if $S_\epsilon$ percolated.
\item In Fig.\ref{rings} there is 
no indication of the ratio going to 0 for increasing $L$. Moreover the
dramatic change in slope around $c=.4$ indicates that the polar cap
$P_{\epsilon'}$ starts forming rings at a smaller value of $c$ than 
the equatorial strip $S_\epsilon$.
\end{itemize}

\begin{figure}[h]
\centerline{\epsfxsize=8.0cm\epsfbox{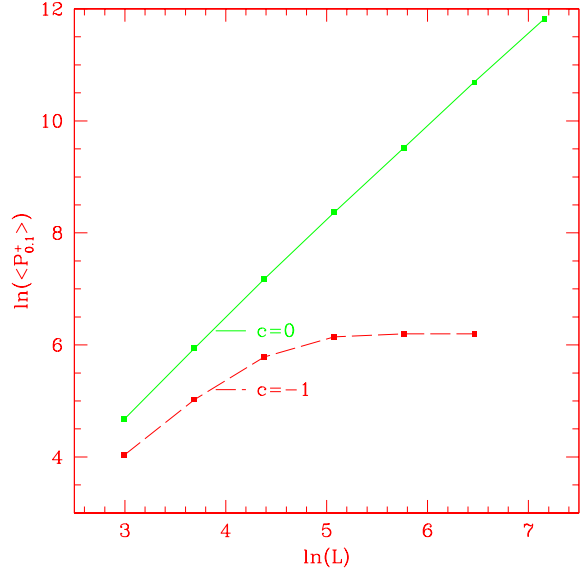}}
\caption{Mean size of clusters of the polar cap $P^+_0.1$ at $c=-1$
(Bernoulli) and at $c=0$ versus $L$. The correlation length $\xi$
is approximately 53 lattice units.}
\label{power}
\end{figure}

We have additional numerical evidence that clusters of a polar cap
$P^+_\epsilon$ smaller than a hemisphere ($s\cdot n>\epsilon/2>0$) form
rings for some $c<1$. Namely we investigated the case $\epsilon=0.1$. For the 
the case $c=-1$ (Bernoulli percolation) it is known rigorously that
clusters of this set have finite mean size. As can be seen from 
Fig.\ref{power} our numerical values at $c=-1$ corroborate this fact.
In the same figure we show the mean cluster size of cluster of
$P^+_{0.1}$ at $c=0$, where the correlation length is approximately 53
lattice units. Even though we increased $L$ up to 1280, the mean 
cluster size shows no sign of leveling off, growing in fact like some
power of $L$, consistent with the formation of rings.

\begin{figure}[htb]
\centerline{\epsfxsize=8.0cm\epsfbox{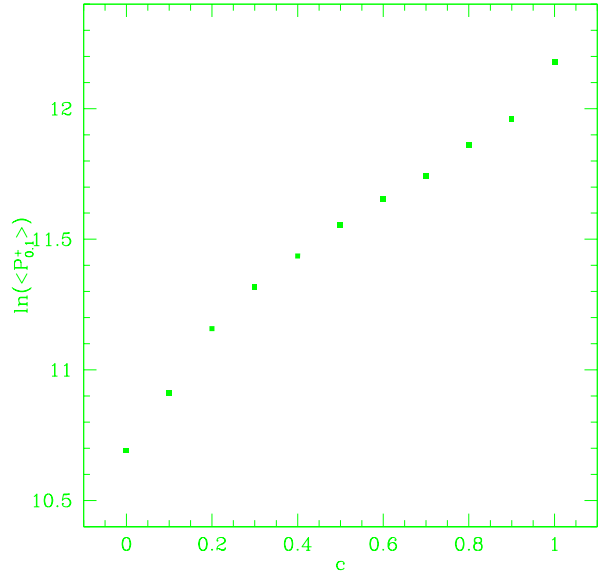}}
\caption{Mean size of clusters of the polar cap $P^+_0.1$ at $L=640$
 versus $c$.} 
\label{monot}
\end{figure}

Therefore there is good numerical evidence that for $c=0$, where we
can reach the thermodynamic limit, clusters of this polar cap form rings.
The natural expectation would be that the mean cluster size of a
subset of a hemisphere is a nondecreasing function of $c$. This is borne
out by the numerics, as shown in Fig.\ref{monot}. There we represent
the mean cluster size of $P^+_{0.1}$ at fixed $L=640$ function of $c$.
The data support the assertion that for any $c>0$ the mean size of
the clusters of $P^+_{0.1}$ diverges, which, via our argument, implies
that the c.a. model must cease being massive for some $c<1$.

Thus we doubt very much that the effects we are seeing represent small
volume artefacts. Moreover, if $s_z$, the $z$-component of the spin
$s$ remained massive at low temperature and in fact an arbitrarily
narrow equatorial strip percolated, one would have to explain away our 
old paradox \cite{ap,npb}: if such a narrow strip percolated, an even 
larger strip would percolate and on it one would have an induced $O(2)$ 
model in its massless phase, in contradiction to the Mermin-Wagner 
theorem.

Consequently it seems unavoidable to conclude that the phase diagram in 
Fig.\ref{phase} represents the truth, that a soft phase exists both in 
the s.a. and the c.a. model and that the massive continuum limit of the 
$O(3)$ model is not asymptotically free. In a previous paper \cite{super}
we have already shown that in nonabelian models even at short distances 
perturbation theory produces ambiguous answers. The present result 
sharpens that result by eliminating the possibility of asymptotic freedom
in the massive continuum limit.

Acknowledgement: AP is grateful to the Humboldt Foundation for a
Senior US Scientist Award and to the Werner-Heisenberg-Institut for its
hospitality.

\end{document}